\documentclass[conference,10pt]{IEEEtran}
\IEEEoverridecommandlockouts
\usepackage{cite}
\usepackage{amsmath,amssymb,amsfonts}
\usepackage{algorithmic}
\usepackage{graphicx}
\usepackage{textcomp}
\usepackage{xcolor}
\usepackage{flushend}
\usepackage{algorithm}
\usepackage{algorithmic}

\def\BibTeX{{\rm B\kern-.05em{\sc i\kern-.025em b}\kern-.08em
    T\kern-.1667em\lower.7ex\hbox{E}\kern-.125emX}}
\begin{document}

\title{Estimating Socioeconomic Status \\ via Temporal-Spatial Mobility Analysis \\
-- A Case Study of Smart Card Data}

\author{\IEEEauthorblockN{Shichang Ding}
\IEEEauthorblockA{\textit{Institute of Computer Science} \\
\textit{University of Goettingen}\\
Goettingen, Germany \\
sding@cs.uni-goettingen.de}
\and
\IEEEauthorblockN{Hong Huang}
\IEEEauthorblockA{\textit{School of Computer Science and Technology} \\
\textit{Huazhong University of Science and Technology}\\
Wuhan, China \\
honghuang@hust.edu.cn}
\and
\IEEEauthorblockN{Tao Zhao, Xiaoming Fu}
\IEEEauthorblockA{\textit{Institute of Computer Science} \\
\textit{University of Goettingen}\\
Goettingen, Germany \\
\{tao.zhao,fu\}@cs.uni-goettingen.de}
}

\maketitle

\begin{abstract}
The notion of socioeconomic status (SES) of a person or family reflects the corresponding entity's social and economic rank in society. Such information may help applications like bank loaning decisions and provide measurable inputs for related studies like social stratification, social welfare and business planning. Traditionally, estimating SES for a large population is performed by national statistical institutes through a large number of household interviews, which is highly expensive and time-consuming. Recently researchers try to estimate SES from data sources like mobile phone call records and online social network platforms, which is much cheaper and faster. Instead of relying on these data about users' cyberspace behaviors, various alternative data sources on real-world users' behavior such as mobility may offer new insights for SES estimation. In this paper, we leverage Smart Card Data (SCD) for public transport systems which records the temporal and spatial mobility behavior of a large population of users. More specifically, we develop S2S, a deep learning based approach for estimating people's SES based on their SCD. Essentially, S2S models two types of SES-related features, namely the temporal-sequential feature and general statistical feature, and leverages deep learning for SES estimation. We evaluate our approach in an actual dataset, Shanghai SCD, which involves millions of users. The proposed model clearly outperforms several state-of-art methods in terms of various evaluation metrics.
\end{abstract}


\section{Introduction}
Socioeconomic Status (SES) is an economically and sociologically combined overall measure of an individual or family, typically based on income level, education level, and occupation~\cite{bradley2002socioeconomic,sirin2005socioeconomic}. SES can be seen as one's economic and social position in relation to others and typically divided into three levels (high, middle, and low)\cite{bradley2002socioeconomic}. An individual with a higher SES means he/she earns more, has a better job or higher education than those with a lower SES. SES nowadays plays an important role in many areas like sociology, economics, public administration, and education. It can help governments to design and evaluate social policies, especially for welfare policy. Recently, companies become more and more interested in assessing people's SES because it is a valuable demographic feature to many emerging applications, such as customized marketing, personalized recommendation, and precise advertisement \cite{szopinski2016factors,chen2016context,hung2005personalized,wu2010contact}. Especially, in personal credit rating, SES is an important factor that helps online banks (e.g., Lending Club\footnote{lendingclub.com, one of the largest peer-to-peer lending platform.}) to decide the volume of loans they will lend to an individual \cite{szopinski2016factors}.\\
\indent
Given its importance, various approaches have been developed to measure SES, most of which need to collect at least one kind of the following information: individual income, education or occupation \cite{bradley2002socioeconomic}, typically through real-world contacts with the individuals under investigation. For a large-scale investigation covering millions of people, it is usually conducted through household interviews by National Statistical Institutes. Some researchers or professional investigation companies also try to collect SES information through methods like online questionnaires or telephone surveys. However, most of them can only cover a small group of people. Although traditional methods can get very detailed information, the investigators usually publish regional-level statistics instead of individual SES information (which is much more important to many companies). Also, the time gap between two successive large-scale surveys could be very long, which may even be several years. If companies decide to collect SES by themselves, they find that the cost is unbearable and many citizens are also quite reluctant to expose their real income or job information. Even governments of some developing countries are also facing the same problem \cite{blumenstock2015predicting}.\\
\indent
Due to the prohibitive costs and time required to collect large-scale individual-level SES information, researchers try to estimate individual-level SES using some easily accessible big data sources like mobile phones call records \cite{soto2011prediction,blumenstock2015predicting,almaatouq2016mobile,xu2018human} or online social networks~\cite{preoctiuc2015analysis,preoctiuc2015studying,lampos2016inferring}, Although most existing big data-based methods can only get a rough income level (low, middle, high) of people. they are still valuable to many companies and researcher, owing to their substantially lower cost and time in estimating SES for a large user population. Further, to better support targeted applications it becomes necessary to improve the accuracy of big data-based SES estimation via better algorithms or different data sources with lower costs or privacy concerns. 
This paper attempts to answer the following question: \emph{Can SES be roughly estimated based on human mobility-related data alone?}\\
\indent
Data-based SES estimation methods are actually based on an observation that different SES levels of people may have different lifestyles. Lifestyle depicts typical routine lives of people. Large-scale human mobility data like smart card data (SCD) or online check-in data can act as an approximation for human lifestyle. Previous methods \cite{soto2011prediction,xu2018human,blumenstock2015predicting} based on cellphone discussed some general statistical mobility features. However, these features are simply complemented to specific cellphone features like the numbers of calls and telephone fares. These mobility features may not be enough for organizations (e.g., public transit agencies) which only have human mobility data. In this paper, we study whether we can get a satisfactory estimation of user-level SES when we only get users' mobility data.\\
\indent
As a case of mobility data source, we take SCD generated by smart card automated fare collection systems, which are now widely used by public transit agencies. Essentially, SCD is administrated by a city municipality and records a large number of individual-level, time-stamped and geo-tagged trip data of its citizens~\cite{bagchi2005potential,mohamed2017clustering}. Although a large and growing body of work has studied SCD in different contexts, little attention has been paid to estimate SES based on SCD. We develop S2S (\emph{Smartcard} to \emph{SES}), a method for estimating SES based on SCD and other related public information. The main challenges in designing S2S are:
\begin{itemize}
\item Designing effective features related to SES based on smart card data.
\item Designing a model which can utilize different types of features to improve the performance of estimation.
\end{itemize}
\indent
To the best of our knowledge, this paper is the first attempt to estimate user-level SES using SCD data. Our main contribution is summarized as follows.
\begin{itemize}
\item We propose a deep neural network (DNN)-based learning approach (S2S), which considers both temporal-sequential features and general statistical features of human mobility. Especially, the sequential aspects are considered in S2S, representing more salient nature of an individual's behavior in socioeconomic context than traditional general statistical features.
\item We evaluated our approach using actual large-scale SCD data of totally 7,919,137 cards of Shanghai City for 16 consecutive days. The results demonstrate our approach significantly outperforms several baselines.
\end{itemize}
\indent
The rest of this paper is structured as follows: related work is reviewed in Section \uppercase\expandafter{\romannumeral2}. Section \uppercase\expandafter{\romannumeral3} introduces the datasets. Section \uppercase\expandafter{\romannumeral4} discusses the features. The S2S model is proposed in Section \uppercase\expandafter{\romannumeral5}. Experimental results on Shanghai SCD are presented in Section \uppercase\expandafter{\romannumeral6}. The paper is concluded in Section \uppercase\expandafter{\romannumeral7} with a brief discussion of limitations and directions of future research.

\section{Related Work}
SES is a widely studied concept in the field of social sciences, especially in health and education analysis \cite{bradley2002socioeconomic}. In recent years, companies and researchers pay increasing attention to SES estimation because of its potential in numerous high-value applications like the personalized recommendation and online banking. Though there has been a great improvement in estimating other demographic attributes like age, ethnicity and gender \cite{zhong2015you,antipov2016minimalistic}, SES estimation still needs more effort. One of the main obstacles is that SES ground truth data (covering a large group of people) is much harder to get than attributes like age and gender. Normally users are more reluctant to disclose their education, occupation and income information. The organizations, which have such data, also seldom open it to the public for privacy reasons. Recently, researchers begin to use indirect SES indicators from some big data sources. These data sources may cover millions of people, recording different aspects of their lifestyles.
\subsection{SES Estimation based on Mobile Phone}
One important data type is mobile phone data. \cite{soto2011prediction} shows that information derived from the aggregated use of cell phone records can be used to identify the socioeconomic levels of a population.  \cite{frias2013cell} provides an analytical model to formalize the relationship between cell phone usage (including mobile phone consumption, social information, and mobility patterns) and socioeconomic indicators (including income and education). \cite{blumenstock2015predicting} estimates Rwandans' SES based on their mobile phone usage. They design a composite wealth index for Rwandans based on whether they have refrigerator, electricity, television and other belongings. Then they extract features from the mobile phone data. The experiments show that the distribution of wealth estimated from mobile phone data has a strong correlation with the distribution of wealth measured by the Rwandan government. This paper considers multiple factors of phone usage including communication, the structure of contact network and mobility pattern. Different from them, we mainly rely on mobility features and use a different kind of data source (SCD). \cite{almaatouq2016mobile} constructs a simple model to produce an accurate reconstruction of district-level unemployment from people's mobile communication patterns alone. \cite{xu2018human} analyses the relationship between multiple mobility features and SES based on mobile phone datasets of two cities: Singapore and Boston. In Singapore, they take the housing price of living area as SES. In Boston, they use the census tracts as SES.They find that the relationship between mobility and SES could vary among cities, and such a relationship is quite complicated. It may be influenced by several different factors like spatial arrangement of housing, employment opportunities, and human activities. For example, phone user groups that are generally richer tend to travel shorter in Singapore but longer in Boston. Our work is different from \cite{xu2018human} in the following ways: 1) we examine the extent to which SES can be estimated from SCD, while they try to figure out the relationships between SES and mobile phone mobility data; 2) we mainly focus on SCD instead of mobile phone; 3) besides the living area, we also consider the work area while labeling people's SES.
\subsection{SES Estimation based on Social Network}
The social network is another important data source which Researchers pay a lot of attention to. \cite{preoctiuc2015analysis,preoctiuc2015studying,lampos2016inferring} all explore how to estimate people's SES based on their tweets. They use the job information from the users' profile as ground-truth. \cite{preoctiuc2015studying} use features like topics, emotions to estimate peoples' income. Their predictions reach a correlation of 0.633 with actual user income, showing that tweets can be used to predict income. \cite{preoctiuc2015studying,lampos2016inferring} further improve the features and significantly increase the accuracy. \cite{huang2016activity} analyzes the relationship between SES and people's activity patterns extracted from Twitter. They find out that while SES is highly important, the urban spatial structure also plays a critical role in affecting the activity patterns of users in different communities.
\subsection{Relationship Study between SES and SCD}
Although SCD records mobility characteristics of a great number of people, the work about the relationship between SES and SCD-based mobility is quite limited. \cite{goulet2016inferring} represents each passenger through a sequence of activities (purely inferred from SCD records) and cluster them using k-means. They survey a part of users about their demographic attributes and then analyze the demographic attributes of each cluster. They find that the average income of some clusters is high than other clusters. So it indicates that income may be related to people's smart card records. \cite{mohamed2017clustering} introduces an approach to cluster passengers living in Rennes (France) based on their temporal habits. They study how fare type proportions are distributed in different clusters. The Rennes SCD dataset includes fare types like Young subscribers, Regular subscribers, Elderly subscribers and etc. They find out there are some mobility differences between different fare type categories. For example, the clusters mainly consisting of students who tend to get back home early in Wednesday since course hours on Wednesdays end early in France, while other clusters do not have this pattern. This also indicates SCD records may be related to users' age and occupation. These works show there is some possible relationship between SCD-based mobility and SES. In this paper, we aim to explore whether and how SCD can be used to estimate SES.
\section{Datasets}
\subsection{Data Collection}
We exploit three related datasets in this paper: smart card, POI and housing price. We describe them respectively below.\\
\indent
\textbf{Smart card:} The smart card dataset is opened by the Shanghai Open Data Applications contest. The dataset contains all the subway records in Shanghai between April 1st and April 16th, 2015. The example format of a subway record is shown in table 1. One single subway trip consists of two successive records. The first one is created when the user gets into the boarding subway station and begin to travel in the subway system. The second record is created when the user gets out of alighting station. If the fare is 0.0, then the user is getting aboard a metro train, or they are getting off. There are 7,919,137 IDs which can be correctly recognized after data cleaning. When users apply for a smart card in Shanghai, they do not need to provide any personal information. So IDs do not have any relationship with real-world identification, avoiding possible privacy leakage.\\
\begin{table}
  \caption{Subway Record Example}
  \label{tab:freq}
  \begin{tabular}{ c | c | p{1.5cm} | c| c}
    \hline
    ID & Date& Time & Station Name & Fare\\
    \hline
    1000019&2015/04/02&17:01:05&station A&0.0\\
    1000019&2015/04/02&17:35:49&station B&4.0\\
    1000039&2015/04/06&18:03:04&station C&0.0\\
    1000039&2015/04/06&18:17:49&station D&2.0\\
    \hline
\end{tabular}
\end{table}
\begin{figure*}
\begin{minipage}{0.3\textwidth}
\centering
\includegraphics[width=2in]{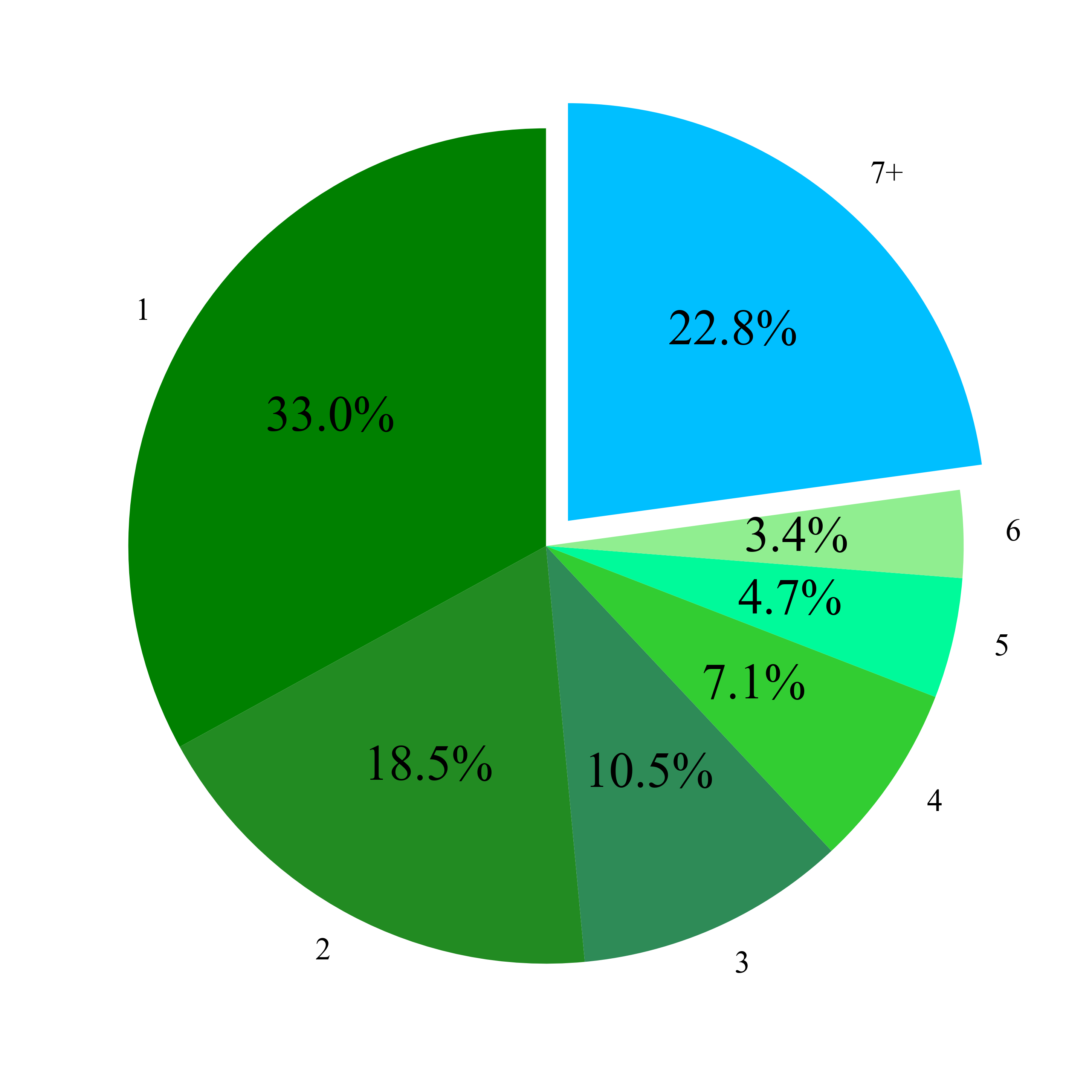}
\caption{User distribution: only 22.8\% are frequent user who take subway more than 7 days}
\label{1}
\end{minipage}
\hfill
\begin{minipage}{0.3\textwidth}
\centering
\includegraphics[width=2in]{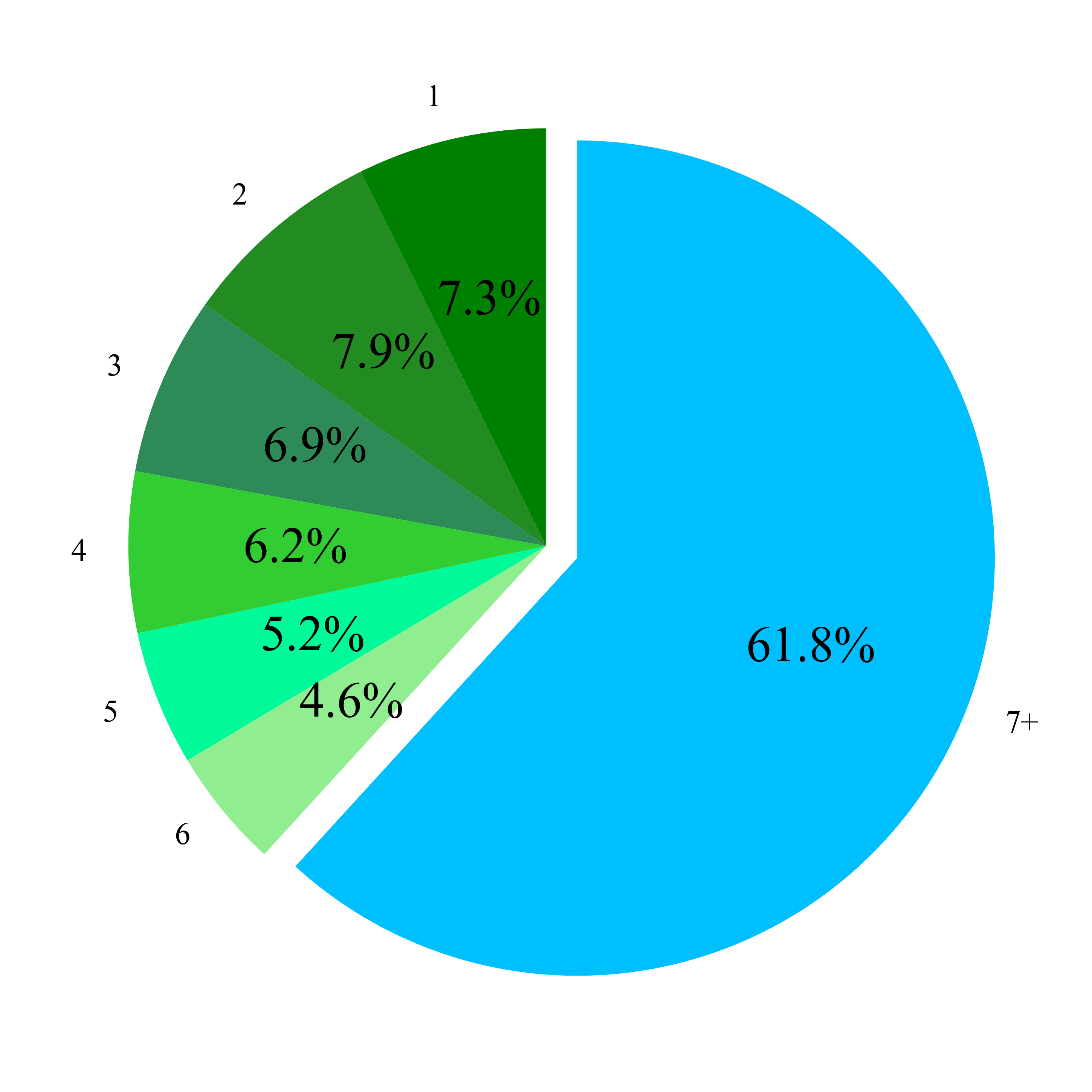}
\caption{Trip Distribution: frequent user takes more than 60\% subway trips}
\label{2}
\end{minipage}
\hfill
\begin{minipage}{0.3\textwidth}
\centering
\includegraphics[width=2.3in]{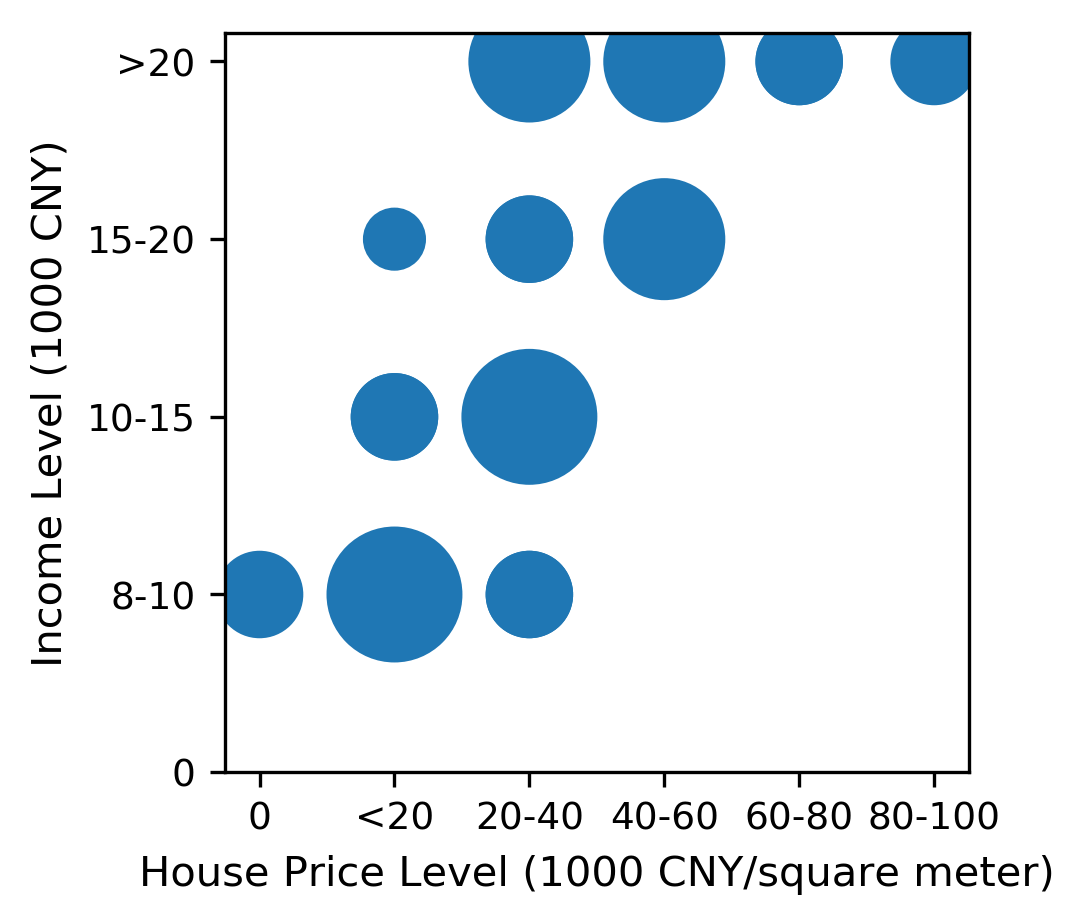}
\caption{The relationship between house price and monthly income: larger size means more people.}
\label{3}
\end{minipage}
\end{figure*}
\indent
\textbf{POI: }POI dataset of Shanghai is crawled based on GaoDe Map API Service\footnote{lbs.amap.com, one of the major online map providers in China}. The categories include Public Facility, Domestic services, Education, Business Residence, Hospital, Hotel, Car services, Sport\&Leisure, Scenery, Restaurant, Public Transportation and Financial Services.\\
\indent
\textbf{Housing price:} Housing price dataset is crawled from Lianjia.com \footnote{sh.lianjia.com, one of the biggest real estate agency service providers in China.} website, which records the house prices and location information of most apartments/houses for selling in Shanghai. We crawl the average housing prices of all communities (a community usually includes many similar houses in one area). 
There are 1,804 communities. The cheapest one is 10,453 CNY/m$^2$. The most expensive one is 99,941 CNY/m$^2$.
\subsection{Ground Truth Construction}
There are two problems in Ground Truth Construction.\\
\indent
First, some users may only use subway for very few times (1 time) during all 16 days. We need to filter users with too less records.\\
\indent
Second, there is no SES information for millions of smart card holders. Automated fare collection (AFC) systems are just designed for billing purpose, so they do not collect socio-demographic information of the card holders in most cities. This Shanghai dataset (appr. 8 million smart cards) is also totally anonymous without any SES-related information, such as occupation, education and income.
And we cannot manually relate smart card IDs with volunteer users because IDs have been hashed before opened for researchers. So it is hard to get actual SES label for each ID. We need to find a reasonable SES label for millions of users.
\subsubsection{Selecting Frequent Users}
As shown in Fig 1, although there are millions of subway users, most of them take very few subway trips.  the largest group of users (33.04\%) only takes subway in 1 day.
More than half of people takes subway in less than 2 days. Only 22.8\% of users have subway trips more than 7 days. And we also checked the trip numbers, 36.9\% of users only took 1 trip. These infrequent users just use the subway occasionally. Subway is not an important transportation method for them. Their mobility data in subway system may be just a random and unimportant action in their regular life. In this paper, we focus on users who have taken subways for at least 7 days. In this way, we selected about 700 thousands frequent users.\\
\indent
Though the number of frequent users is much smaller than infrequent users, the total number of trips they take is much more than the others. As shown in Fig 2, more than 60.1\% of trips are taken by frequent users, who take subway more than 7 days.
\subsubsection{Labeling Frequent Users}
Getting SES label is a common problem when estimating SES for a large number of people\cite{blumenstock2015predicting,smith2016beyond,filmer2001estimating}. Many works use the housing price of people's living place as a proxy to represent people's possible SES \cite{mohamed2017clustering,xu2018human,huang2018cross,harris2014application,juhn2011development,ghawi2015novel,coffee2013relative}. And \cite{xu2018human} finds out that the average housing price and the income level at the corresponding area are strongly correlated (0.88). As shown in Fig. 3, we also held an online survey \footnote{http://wj.qq.com/s2/3598293/4053/}, which collect 78 Shanghai inhabitants' monthly income and housing price. To protect the privacy and get more successful responses, we use income levels (e.g, 5,000-10,000 CNY) instead of accurate numbers. So some answers may overlap in Fig. 3. We use the size of the bubble to show the overlapped number. Bigger bubble means more same answers. We can see, the income level generally increases along with the housing price. Pearson's correlation is 0.68. The correlation is not so strong as in \cite{xu2018human}. This may be partially caused by the phenomenon in China that some low-income young people buy high-priced houses with the help of their families. However, high family income may still also be a ``bonus" to people's SES. So in general, we think housing price is a good indicator of the people' SES.\\
\indent
In this paper, we use people's house price as an approximation of frequent users' SES. 
First, we use the method in \cite{zhou2014commuting} to find frequent users' home station (the station nearest to their home). Then, we select the communities around the home stations (less than 2 km), to calculate the average housing price of the home station. SES is usually divided into 3 levels: high, middle and low. We divide frequent users into 3 levels based on the average housing price of their home station. There are 19.4\% of users at high level (housing price $>$ 70000 CNY/m$^2$), 36.2\% in middle level and 44.4\% at low level.
\section{Feature Engineering}
\subsection{Overview}
A user's smart card records can be seen as a list of tuples $\{(s_{1}, t_{1},ao_{1} ), (s_{2}, t_{2},ao_{2}), \ldots, (s_{n}, t_{n},ao_{n})\}$. $s_{i}$ and $t_{i}$ denote the subway station and the time of the $i$-th record. $ao_{i}$ denotes whether the user is getting aboard and off at $i$-th record. Given users' smart card records, we aim to estimate users' SES levels. The overall research design is shown in Fig. 4. One of the key challenges is feature engineering. We mainly utilize two types of features in this paper: general statistical features and temporal-sequential feature. General features (shorten form of general statistical features) usually consider the statistical features of a user's whole mobility data. They have been discussed by previous works like \cite{soto2011prediction,pappalardo2015using,xu2018human}. However, previous papers largely neglect the temporal and function information related to each station, which will be discussed in following section.
\subsection{General Feature}
\subsubsection{$F_{rg}$, Radius of Gyration } $F_{rg}$ is defined as follows:
\begin{equation}
F_{rg}=\frac{\sum_{i=1}^{n}distance(\overrightarrow{s_{i}},\overrightarrow{s_{c}})}{n}
\end{equation}
\indent
Here, $\overrightarrow{s_{i}}$⃗ denotes the location (latitude and longitude coordinates) of $s_{i}$. $\overrightarrow{s_{c}}=\sum \overrightarrow{s_{i}} / n$ denotes the geographic center of all $s_{i}$. $distance$ is the geographic distance between two locations. A large value of $F_{rg}$ indicates the user mobilize in a large area.\\
\subsubsection{$F_{krg}$, K-Radius of Gyration} Let $count(i)$ be a counting function, which is equal to the number of $s_{i}$ in a user's whole mobility record. A large value of $count(i)$ means the user often visit the subway station $s_{i}$. $F_{krg}$ is a radius of gyration calculated using only top $k$ visited stations. \cite{pappalardo2015using} proposed it to measure how a user's top $k$ stations determine his/her radius of gyration. $F_{krg}$ is defined as:
\begin{equation}
F_{krg}={\frac{\sum_{i=1}^{k}(count(i)  \cdot distance(\overrightarrow{s_{i}},\overrightarrow{s_{c}}))}{\sum_{i=1}^{k} count(i)}}
\end{equation}
The aim of $F_{krg}$ is to find out returners and explorers. \cite{pappalardo2015using} suggested that, k-returners are those whose $F_{krg} \geq F_{rg} / 2$ and k-explorers are those for whom $F_{krg} < F_{rg} / 2$. We can simply think that k-returners are those who tend to spend most of the time between k the most important locations, while k-explorers are those whose activity space cannot be well described by only $k$ top locations. And in this paper, we set $k=2$. In this way, 2-returners are likely to be a common commuter between home and working place.
\subsubsection{$F_{nds}$, Number of Different Stations} $F_{nds}$ is defined as follows: 
\begin{equation}
F_{nds}=|{set}(s_{1}, s_{2}, \ldots, s_{n})|
\end{equation}
\indent
$F_{nds}$ measures the total number of different stations visited by a user during all 16 days. A larger value of $F_{nds}$ means that the users tend to visit more different subway stations.
\begin{figure}
\centering
\includegraphics[width=3.5in]{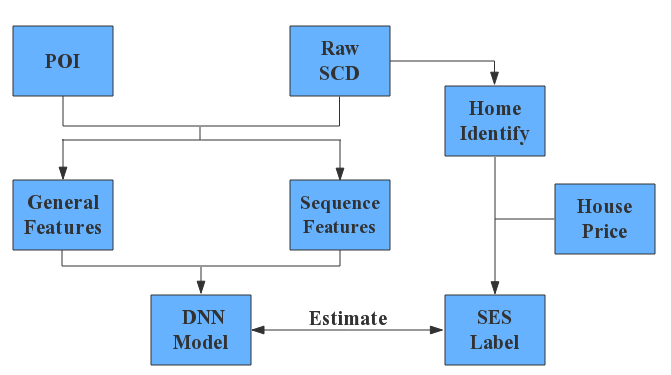}
\caption{Overall research design}
\end{figure}
\subsubsection{$F_{ae}$, Activity Entropy} 
Given a vector $\{p_{1}, p_{2}, \dots, p_{\overline{n}}\}$, where $\overline{n} = F_{nds}$ and $p_{i}=\frac{count(i)}{\sum_{i=1}^{n} count(i)}$. $p_{i}$ denotes the proportion of visiting numbers of station $s_{i}$, the activity entropy is calculated as:
\begin{equation}
F_{ae}=-\sum_{i=1}^{\overline{n}} p_{i} \log (p_{i})
\end{equation}
A large value of $F_{ae}$ means that the spatial diversity of a user's daily activities is high.
\subsubsection{$F_{td}$, Travel Diversity} Travel diversity measures the regularity of a user's movements among his/her subway stations. We define an origin-destination trips as a trip between two consecutive stations. Let E denote all the possible origin-destination pairs (without considering direction) extracted from $\operatorname{set}\left(s_{1}, s_{2}, \ldots, s_{\overline{n}}\right) )$, all stations a user visit. Then the travel diversity is defined as:\\
\begin{equation}
F_{td}=-\sum_{i \in E} p_{i}^{\prime} \log (p_{i}^{\prime})
\end{equation}
where $p_{i}^{\prime}$ is the probability of observing a trip between the $i$-th origin-destination pair. A large value of $F_{td}$ means that a user's tend to travel between quite different origin stations and destination stations.

\subsection{Sequence Feature}
People may tend to follow regular and stable patterns during their everyday lives. And people in different SES-level may visit different places and have different commute schedules. For example, cleaners usually need to go to company earlier while IT engineers may have to work at company until very late at night. Here we use sequence feature (shorten form of temporal-sequential feature) to describe these phenomenons.\\
\indent
We divide all 16 days into 1536 (16x24x4) time bins by every 15 minutes. For each time bins, we need to find the location where a user stay, and calculate a feature vector based on the location. Given that a user's sequence feature is $\{X_{1},X_{2},\ldots,X_{i},\ldots,X_{N}\}$, where $N=1536$ and $X_{i}$ denote the feature vectors of location at the $i$-th time bins. $X_{i}$ consists of three kinds of features: the ID of time bins ($timeID$, from 0 to 1535), function of station for most citizens ($F_{fm}$,$\{residential, entertainment, working, transfer\}$) and function of station for current user ($F_{fu}$,$\{home, work, others, transfer\}$).\\
\indent
To find the location where a user stay, first we take the stations as the location of the corresponding time bins. For example, if during the first time bins, a user get aboard on station A, then we take station A as the user's location of the first time bins.\\
\indent
Then for time bins which there is no corresponding station, we use following method to find their approximate locations:\\
\indent
\textcircled{\small{1}} Among the time bins with a station location, find out those when the user is getting aboard and the others when the user is getting off, based on $ao_{i}$. The former time bins are denoted as $T_{aboard} = \{t_{a1},t_{a2},\ldots,t_{ai}\ldots\}$. The latter time bins are denoted as $T_{off} = \{t_{o1},t_{o2},\ldots,t_{oi}\ldots\}$.\\
\indent
\textcircled{\small{2}} If a series of time bins are between two consecutive stations, $t_{oj}$ and $t_{ak}$( the first for getting off and the second for getting aboard), the locations of the first half time bins are the station of $t_{aj}$ while the second half are the station of $t_{ak}$.\\
\indent
\textcircled{\small{3}} If a series of time bins are between two consecutive stations, $t_{al}$ and $t_{om}$( the first for getting aboard and the second for getting off), we do not need to find their locations. The detail of how to calculate the feature vectors for these time bins will be discussed in following sections.\\
\indent
\textcircled{\small{4}} For the time bins before $t_{a1}$, the locations are the station of $t_{a1}$.\\
\indent
\textcircled{\small{5}} For the time bins after last getting off station (i.e, $t_{oN}$), the locations are the station of $t_{oN}$.
\subsubsection{$F_{fm}$, Function of station for Most citizens}
The step of urbanization leads to different functional regions in a city, e.g., residential areas, business districts, and entertainment areas \cite{yuan2015discovering}. People show in the different functional areas may have different social attributes. For example, housewives may mainly stay inside residential areas while regular office worker may travel between the residential area and business districts during the weekday. And different kinds of people may spend different time in some special functional regions. For example, a rich family may spend more time in entertainment areas during the weekend than an ordinary family. Here we use two features called $F_{fm}$ to describe this phenomenon.\\
\indent
Here we explain how to determine the function for each subway station. There are different functional regions in one city, supporting different needs of people's urban lives. And similarly, each subway station also has a different function. People tend to use the subway station which is nearest to their starting location and ending location. For example, if a subway station is inside a residential area, then most people using this subway should be the people who live near this station. During the weekday, most users of this subway station would get into the subway in the morning to go to work and get out of the station in the evening to go back home. On the other hand, if a subway station is inside a work area, surrounded by a lot of companies, then most people using this subway should be the people who work near this station. During the weekday, most users of this subway station would get out the subway in the morning to go to work and get into the station in the evening to go back home. So the function of one subway station is actually the function of the area near it.\\
\indent
\begin{figure}
\includegraphics[width=3.4in]{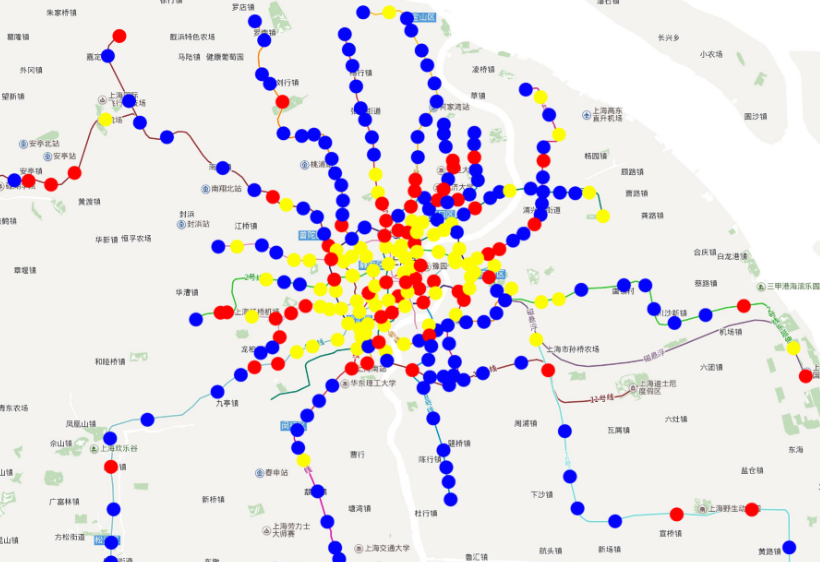}
\caption{Function Station Distribution in Shanghai: blue = residential, red = entertainment, yellow = work, the lines are the subway lines, the points are the subway stations.}
\end{figure}
In this paper, we use the same method in \cite{yuan2015discovering} to divide all Shanghai subway station into 3 kinds: residential, entertainment and work. This method needs to consider the human mobility and poi data of each station. The distribution of function stations is shown in Fig. 5. The blue points represent residential stations, the red points represent entertainment stations and the yellow points represent work stations.\\
\indent
For most $X_{i}$, $F_{fm}$ is ``residential", ``entertainment" or ``working". However, if $X_{i}$ is between two consecutive stations, $t_{al}$ and $t_{om}$( the first for getting aboard and the second for getting off), $F_{fm}$ is ``transfer". It means the user is traveling from one function area to another function area.
\subsubsection{$F_{fu}$, Function of station for current User}
For some users, the function of a specific station may be different from most users. For example, someone may work in a supermarket in a living area. Though for most people, the station is a ``residential" station. However, for this person, the station is more like a ``working" station.\\
\indent
In this paper, we use the same method in \cite{yuan2015discovering} to divide a user's stations into 3 kinds: ``home", ``work" and ``others". For most $X_{i}$, $F_{fu}$ is ``home", ``work" or ``others". However, if $X_{i}$ is between two consecutive stations, $t_{al}$ and $t_{om}$( the first for getting aboard and the second for getting off), $F_{fu}$ is ``transfer".
\section{S2S Model}
The goal of the proposed model is to estimate a user's SES level, denoted as $\boldsymbol{Y}_{uid}$, where $uid$ is the id of a smart card user. Fig. 6 shows the architecture of the proposed model, which is comprised of two major components. The sequential component processes sequence features and outputs $\boldsymbol{Y}_{s}$. The general component processes general feature and outputs $\boldsymbol{Y}_{g}$. $\boldsymbol{Y}_{s}$ and $\boldsymbol{Y}_{g}$ are fused and fed into the softmax layer to estimate the SES level of input user.
\subsection{Sequential Component}
People of different SES level may have different lifestyles, like visiting different places and having different commute schedules. We need to capture the temporal dependence of people's mobility. The recurrent neural network (RNN) is an artificial neural network which is widely used for capturing the temporal dependency in sequential learning, such as the natural language processing and speech recognition \cite{he2016deep}. When processing the current time step in the sequence, it updates its memory (also called hidden state) according to the current input and the previous hidden state. The output of the recurrent neural network is the hidden state sequence at all the time steps in the sequence. The sequential feature we design considered the transition of different function stations, which can be effectively handled by RNN. Sequential component is composed of an embedding layer, a single RNN layer, and two fully-connected layers, as shown in Fig. 6. In this paper, we denote the feature at time bin $i$ as $\boldsymbol{X}_{i}=(timeID, F_{fm}, F_{fu})$. In our experiments, RNN performs not so well in processing the long time bins due to vanishing gradient and exploding gradient problems. Therefore, instead of the RNN layer, we adopt the Long Short-Term Memory (LSTM) \cite{gers1999learning} layers. In short, LSTM adds an input gate and a forget gate to alleviate the gradient vanishing/exploding problem.\\
\indent
$\boldsymbol{X}_{i}$ is fed into an embedding layer first. Because $timeID$, $F_{fm}$ and $F_{fu}$ are both categorical values which can not feed to RNN layer
directly \cite{gal2016theoretically}. The embedding layer transform $timeID$, $F_{fm}$ and $F_{fu}$ into three low-dimensional real vectors ($timeID^{e}$, $F_{fm}^{e}$ and $F_{fu}^{e}$ ), respectively. The $timeID^{e}$, $F_{fm}^{e}$ and $F_{fu}^{e}$ are concatenated to get  $\boldsymbol{X}^{e}_{i}$. $\boldsymbol{X}^{e}_{i}$ is fed into the LSTM layer, which output a hidden state $\boldsymbol{h}_{i}$. We concatenate all of the hidden state fragment $[\boldsymbol{h}_{1}, \boldsymbol{h}_{2}, \ldots, \boldsymbol{h}_{i},\ldots,\boldsymbol{h}_{N}]$ as ${\boldsymbol{H}_{N}}$. Then ${\boldsymbol{H}_{N}}$ is fed into the fully-connected layer as:
\begin{equation}
\boldsymbol{H}_{s}=ReLU(\boldsymbol{W}_{hs}\boldsymbol{H}_{N}+\boldsymbol{b}_{hs})
\end{equation}
$\boldsymbol{H}_{s}$ is then fed through the fully-connected layer to output the $\boldsymbol{Y}_{s}$, defined as:
\begin{equation}
\boldsymbol{Y}_{s}=\tanh (\boldsymbol{W}_{s} \boldsymbol{H}_{s}+\boldsymbol{b}_{s})
\end{equation} 
where $\boldsymbol{W}_{hs}, \boldsymbol{b}_{hs}, \boldsymbol{W}_{s}$ and $\boldsymbol{b}_{s}$ are the learnable parameter matrices used in the fully-connected layers.
\begin{figure}
\includegraphics[width=3.6in]{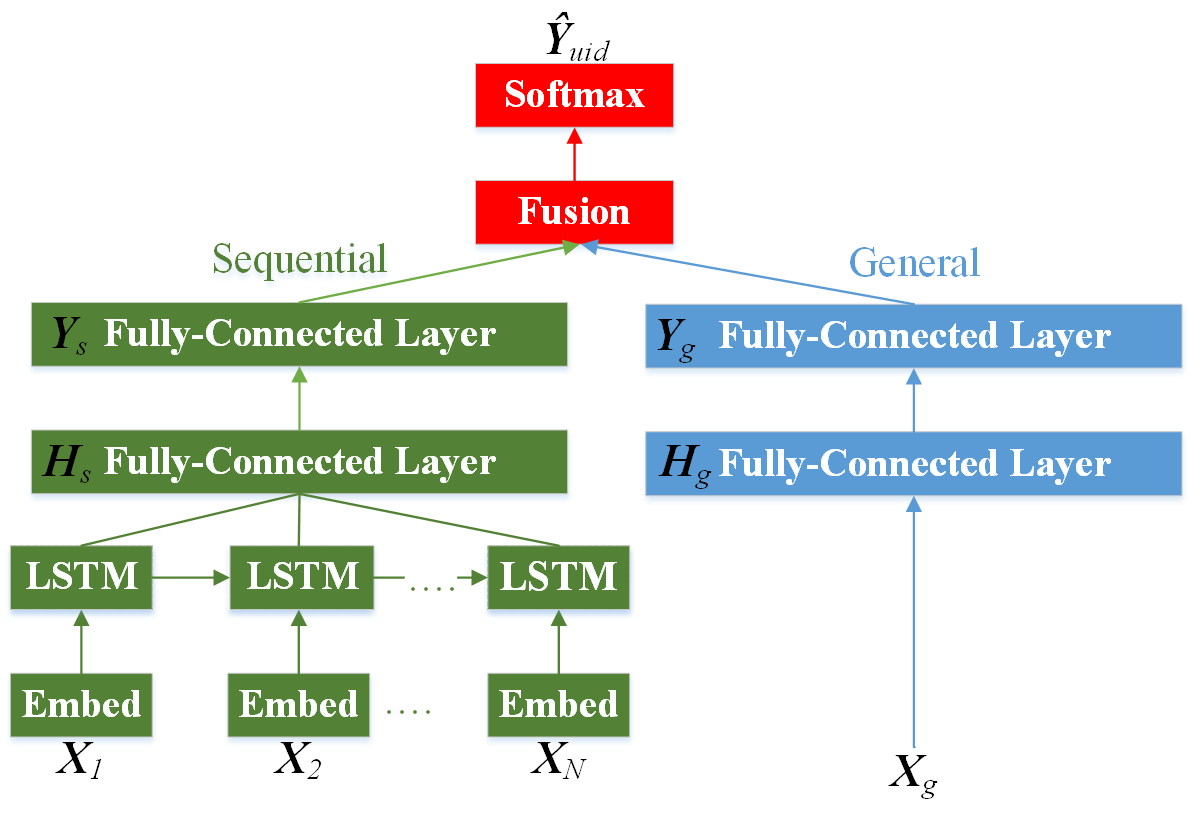}
\caption{Model Architecture}
\end{figure}
\subsection{The Structure of General Component} 
Besides the sequence feature, the general mobility feature may also reflect a part of lifestyles. We discussed these features and their possible relationship with SES-level in Section \uppercase\expandafter{\romannumeral4}. We stack two fully-connected layers to model the general factors that affect SES. $\boldsymbol{X}_{g}= [F_{rg},F_{krg},F_{nds},F_{ae}]$. The first layer processes the feature vector $\boldsymbol{X}_{g}$ and outputs a hidden state $\boldsymbol{H}_{g}$:
\begin{equation}
\boldsymbol{H}_{g}=ReLU(\boldsymbol{W}_{h g} \boldsymbol{X}_{g}+\boldsymbol{b}_{h g})
\end{equation}
\indent
Then $\boldsymbol{H}_{g}$ is fed into the second layer and get the output of the general Component $\boldsymbol{Y}_{g}$:
\begin{equation}
\boldsymbol{Y}_{g}=\tanh(\boldsymbol{W}_{g} \boldsymbol{H}_{g}+\boldsymbol{b}_{g})
\end{equation}
where $\boldsymbol{W}_{h g}, \boldsymbol{b}_{h g}, \boldsymbol{W}_{g}$ and $\boldsymbol{b}_{g}$ are the learnable parameter matrices used in the fully-connected layers.
\subsection{Fusion and Training}
We here combine the output of the two components as shown in Fig. 6. The fusion layer assigns the weights to two components. Finally, the softmax layer estimates the SES level of a user denoted by $\boldsymbol{\hat{Y}}_{uid}$. $\boldsymbol{\hat{Y}}_{uid}$ is defined as:
\begin{equation}
\boldsymbol{\hat{Y}}_{uid}=Softmax(\boldsymbol{V}_{s} \circ \boldsymbol{Y}_{s}+\boldsymbol{V}_{g} \circ \boldsymbol{Y}_{g})
\end{equation}
where o is element-wise multiplication, $V_{s}$ and $V_{g}$ are the learnable parameters that adjust the contribution of sequence and general features to $\boldsymbol{\hat{Y}}_{uid}$. The model can be trained by minimizing the cross-entropy between the ground truth $\boldsymbol{Y}_{uid}$ and the estimated SES level $\boldsymbol{\hat{Y}}_{uid}$:
\begin{equation}
\zeta(\theta)=-\boldsymbol{{Y}}^U_{uid} \log \boldsymbol{\hat{Y}}^U_{uid}
\end{equation}
where $\theta$ are all learnable parameters of S2S model and $U$ means the user number for training. We first construct the training dataset from a part of users' actual SES level and corresponding features. Then, S2S model is trained via back-propagation and Adam \cite{kingma2014adam}.

\section{EXPERIMENTS}
\subsection{Settings}
The details of datasets and ground truth are already introduced in Section \uppercase\expandafter{\romannumeral3}. Finally, We picked 729,859 users who take the subway for at least 7 days (during 16 days). These users are divided into 3 SES levels: high, middle and low. 80\% of picked users are for training and 20\% for testing. The results are mainly measured by classification precision, recall, and F1-score.\\
\indent
To the best of our knowledge, there exists no model directly estimating SES from users' SCD. We use the following baselines to test the effectiveness of our model:\\
\indent
\textcircled{\small{1}} \textbf{Random Guess} just randomly classifies the user to an SES label.\\
\indent
\textcircled{\small{2}} \textbf{STL}. This method predicts twitter users' demographics based on their online check-ins \cite{zhong2015you}. Online check-ins are another kind of mobility data. They are uploaded to online social networks by people to show where and when they are. STL organizes users’ check-ins into a three-way tensor representing features based on spatial, temporal and location information (e.g, location category, keywords, and reviews of a POI). Then a support vector machine (SVM) is trained for estimate users’ demographics (e.g., gender, blood type). We treat station records as users' check-ins when using STL. However, we have to omit some location information like reviews. Because subway stations just do not these kinds of data.\\
\indent
\textcircled{\small{3}} \textbf{Gradient boosting decision tree (GBDT)}. The gradient boosting model is famous for its outstanding performance and efficiency for estimation. The LightGBM is an open source gradient boosting library \cite{ke2017lightgbm}. It has been widely adopted in many data mining competitions like Kaggle. We use sequence feature and general feature to train LightGBM model.\\
\indent
Besides the above baselines, Sequence model (\textbf{S2S-S} model) and General model (\textbf{S2S-G} model) are also tested to find out the most effective feature categories. S2S-S model only uses sequential features with sequential component. S2S-G model only uses general features with general component. We refer our method which involves both sequence and general feature as \textbf{S2S-SG}.\\
\indent
\textbf{Parameter Setting}. The main parameters of our experiment are as follows:
\begin{itemize}
\item In the embedding layer, we embed $timeID$ to $R^{11}$, $F_{fm}$ to $R^2$ and $F_{fu}$ to $R^2$
\item In the general component, the neuron number of two fully-connected layers are both 24 neurons.
\item In the sequential component, the size of the hidden vector $h_{i}$ is 64.
\item In the fusion component, the size of the hidden vector $Y_{s}$ is 24.
\end{itemize}
\indent
The learning rate of Adam is 0.001 and the batch size during training is 12000. Our model is implemented with Keras. We train our model on a 64-bit server with 12 CPU cores, 64GB RAM and NVIDIA 1080Ti GPU with 12G VRAM.
\subsection{Performance Comparison}
Table 2 shows the performance of baselines and S2S, and note the averages of 3 classes are used as the main comparison metric. From the result, we can see that all the metrics of S2S-SG performs better than all baselines, achieving 69\% in precision, 67\% in recall and 68\% in F1-score. Table 3 shows the performance of S2S-SG in each SES class.\\
\indent
As shown in Table 2, STL is clearly better than Random Guess while less accurate than LightGBM. The reason why STL does not perform well on smart card dataset might be caused by two reasons. First, STL did not design features or methods specifically for SES estimation. Also, the subway station does not have one of the important information which STL relies on, i.e., people’ reviews and keywords. Reviews and keywords of locations may also contain useful information about SES. However, unlike restaurants in STL, subway station did not have similar review information.
LightGBM is better than STL, showing the proposed features are more suitable to estimate SES based on SCD. Lightgbm underperforms S2S-GS, likely due to the fact Lightgbm underperforms LSTM on understanding long sequential features.\\
\indent
We can also see that S2S-SG outperforms the other S2S models. S2S-S is clearly better than S2S-G, demonstrating the value of sequential features. And the performance of S2S-S is even better than LightGBM with full features. There may be two reasons why general statistical features are not so useful as sequential features. First, the dataset covers only 16 days. The cellphone datasets which previous works studied usually last for months. So the general feature here may be not suitable for short time. Second, general features are not good at capturing some subtle differences in people's lifestyles. For example, some high SES-level people like to go for entertainment instead of going back home after work, while some low SES-level people also visit such an area for part-time work. It is hard to distinguish them based on general features because they may all have a larger mobility area than others, like home-work commuters. However, sequential features can help in these scenarios, e.g., checking whether one goes to a station for work or for entertainment, or checking whether one is going to an entertainment area during usual working time (e.g, 9am-5pm every workday) or after work (e.g, after 8 pm). Also, people who go to entertainment areas during work time are more likely to be a service staff than a consumer.
\begin{table}
  \centering
  \caption{Comparison of each methods}
  \label{tab:freq}
  \begin{tabular}{c|c|c|c}
    \hline
    Algorithm & Precision & Recall &F1\\
    \hline
    Random Guess & 0.35 &  0.33  &  0.33\\
    \hline
    STL & 0.49 &  0.42  &  0.45 \\
    \hline
    LightGBM & 0.58  & 0.57 & 0.58\\
    \hline
    S2S-S  & 0.63 & 0.62 & 0.63 \\
    \hline
    S2S-G  & 0.53 & 0.51 & 0.52 \\
    \hline
    S2S-SG & 0.69 & 0.67 & 0.68  \\
    \hline
\end{tabular}
\end{table}

\begin{table}
\centering
  \caption{Performance of S2S-SG}
  \label{tab:freq}
  \begin{tabular}{c|c|c|c}
    \hline
    SES-Level & Precision & Recall &F1\\
    \hline
     High   &    0.69  &  0.55   &  0.61      \\
     \hline
     Middle &    0.65  &  0.67   &  0.66      \\
     \hline
     Low  &    0.74  &   0.80  &   0.77     \\
    \hline
    Avg &    0.69  &  0.67   &  0.68      \\
    \hline
\end{tabular}
\end{table}
\indent
We also manually check some error estimations. We find out that many users in high SES-Level are mislabeled as middle SES-level. This may be because most frequent SCD users are not so ``rich". Actually, most subway-frequent users are middle and low-income levels among the city's population, so their difference may not so clear. Besides, we just differ high SES-level or middle SES-level people based on their housing price (70,000 CNY/m$^2$). However, there is a large group of users who are around the 70,000 CNY/m$^2$. We checked their home stations. Many middle and high price-level home stations are quite near to each other. So the difference of mobility feature between them is also not so clear. It means we still need to improve the features in our future work.\\
\section{Conclusion}
This paper examines whether people's SES can be estimated only based on their smart card mobility data. We take the Shanghai smart card data as a case study. Because individual-level income information is hard to get for millions of people, we hypothesize that people's income level is related to the house-price level of their home. In this way, we get the SES label of about 700 thousand users who frequently take subways. Mobility features and a DNN model, S2S, are proposed to estimate their SES-level. In the end, experiments show that these SCD-based features can be used to estimate the SES level (much better than random guess), wherein the sequential features are clearly better than traditional general features. This method can be used to quickly give a rough individual-level SES estimation for millions of people, when companies or researchers can only get people's mobility data.\\
\indent 
This paper is the first try to estimate SES from SCD, validating the predictive power of SCD-based mobility data on SES. There are still problems we need to solve. For example, because we use the house price of people's living area as the ground truth. We cannot leverage some important features (e.g., favorite locations and housing price level of their working area) in estimating SES. We plan to conduct a detailed SES survey of reasonable scale to build a more precise model between SES and mobility as future work.
\section{Acknowledgment}
The research work was partly funded by the European Union’s Horizon 2020 research and innovation program under the Marie Sklodowska-Curie grant agreement No. 824019, National Natural Science Foundation of China (No. 61802140) and Hubei Provincial Natural Science Foundation (No. 2018CFB200).
Also, we would like to acknowledge Xu Wang for her support in the data access and discussions on mobility patterns, and Jar-Der Luo for his insights on socioeconomic status. 

\bibliographystyle{IEEEtran}  
\bibliography{IEEEabrv,myieee}%

\end{document}